\documentclass{iau}
\usepackage{graphicx}

\title[Cosmological Evolution of SMBHs] 
{Cosmological Evolution of Supermassive Black Holes: Mass Functions and Spins}

\author[Li, Wang \& Ho]   
{Yan-Rong Li $^1$, Jian-Min Wang $^1$
 \and  Luis C. Ho $^2$}

\affiliation{$^1$ Key Laboratory for Particle Astrophysics, Institute of High 
Energy Physics, \\ 19B Yuquan Road, 
Beijing 100049, China \\email: {\tt liyanrong@mail.ihep.ac.cn, wangjm@mail.ihep.ac.cn} \\[\affilskip]
$^2$The Observatories of the Carnegie Institution of Washington, \\
813 Santa Barbara St., Pasadena, CA 91101, USA \\email: {\tt lho@obs.carnegiescience.edu}}

\pubyear{2012}
\volume{290}  
\jname{Feeding compact objects: Accretion on all scales}
\editors{C.M. Zhang, T. Belloni, M. M\'endez \& S.N. Zhang, eds.}

\begin{document}

\maketitle

\begin{abstract}
We derive the mass function of supermassive black holes (SMBHs) over the redshift range
$0<z\lesssim2$, 
using the latest deep luminosity and mass functions 
of field galaxies.  Applying this mass function, combined with the bolometric 
luminosity function of active galactic nuclei (AGNs), into the the continuity 
equation of SMBH number density, we explicitly obtain the 
mass-dependent cosmological evolution of the 
radiative efficiency for accretion. 
We suggest that the accretion history of SMBHs and their spins evolve in two distinct 
regimes: an early phase of prolonged accretion, plausibly driven by major 
mergers, during which the black hole spins up, then switching to a period of 
random, episodic accretion, governed by minor mergers and internal secular 
processes, during which the hole spins down.  The transition epoch depends on 
mass, mirroring other evidence for ``cosmic downsizing'' in the AGN 
population.
\end{abstract}

\firstsection 
\section{Introduction}
It has been realized that SMBHs assemble their mass predominantly through accretion; 
however, how SMBHs are fueled remains a unsolved issue. 
The spin of SMBHs traces the angular momentum of the accreted material and therefore
is a powerful cosmic probe of SMBH feeding. 
The link between the radiative efficiency for accretion and black hole spin
allows us to analyze the net angular momentum of the accreted gas by quantifying 
the radiative efficiency. 
The previous study by \cite[Wang et al. (2009)]{Wang09} 
on the cosmological evolution of the radiative
efficiency had shown that SMBHs are spinning down with time since $z\approx2$, 
strongly implying random accretion onto SMBHs.

Starting from the continuity equation for SMBH mass function, 
with the help of the duty cycle of SMBHs $\delta(t, M_\bullet)$ and the 
mean mass accretion rate $\langle\dot M_\bullet\rangle=\delta\dot M_\bullet$, 
we can obtain the the radiative efficiency 
from observables by (see \cite[Li et al. 2012]{Li_etal12} for details)
\begin{equation}
\small
\eta^{-1}(z, M_\bullet)=1+\frac{c^2}{\dot{u}(z,M_\bullet)}
    \left(\frac{{\rm d} t}{{\rm d} z}\right)^{-1} 
    \frac{\partial }{\partial z}
    \int^\infty_{M_\bullet}N(z,M_\bullet'){\rm d}{M_\bullet'},
\label{equ1}
\end{equation}
where $\dot{u}(z,M_\bullet)=\bar\lambda L_{\rm Edd} N_{\rm AGN}$,
 $\bar\lambda$ is the mean Eddington ratio, $L_{\rm Edd}$ is the Eddington luminosity, 
 and $N(z, M_\bullet)$ consists of the SMBH mass functions
 in galaxies ($ N_{\rm G}$) and AGNs ($ N_{\rm AGN}$). This equation is the generalized $\eta-$equation
 of \cite[Wang et al.~(2009)]{Wang09}.
 
\section{SMBH mass functions}
\begin{figure}[h]
\centering
 \includegraphics[angle=-90.0, width=3.6in]{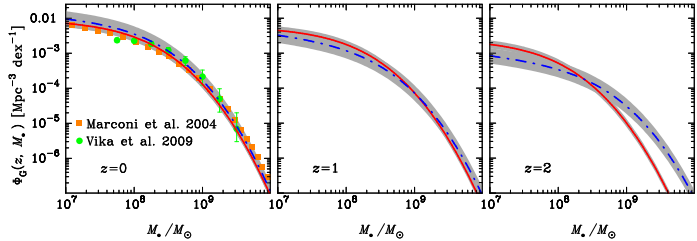} 
 \caption{SMBH mass function in normal galaxies at $z$ = 0, 1, and 2, derived from the galaxy LF 
(red solid lines) and the galaxy SMF (blue dot-dashed lines).}
\label{fig1}
\vspace*{-0.2 cm}   
\end{figure}

We derive the SMBH mass functions in normal galaxies from the galaxy luminosity function 
(LF; \cite[Cirasuolo et~al. 2010]{Cirasiolo10}) and
stellar mass function (SMF; \cite[P{\'e}rez-Gonz{\'a}lez et al. 2008]{Perez-Gonzalez2008}), 
respectively. We use three ingredients: (1) the bulge-to-total luminosity ratio;
(2)~a relation between SMBH  and  spheroid mass;
and, for the method involving the galaxy LF, (3) a prescription to describe the 
passive evolution of the spheroid luminosity.  
Fig. \ref{fig1} shows the obtained SMBH mass functions at $0<z\lesssim2$. 
A good agreement between the two methods can be found (\cite[Li et al 2011]{Li_etal11}).

The SMBH mass function of AGNs is calculated by combining the observed
Eddington ratio distribution and the AGN
bolometric LF of \cite[Hopkins et al. (2007)]{Hopkins_etal2007}.

\section{Evolution of the radiative efficiency and SMBH spins}
\begin{figure}[h]
\centering
 \includegraphics[angle=-90.0, width=5.2in]{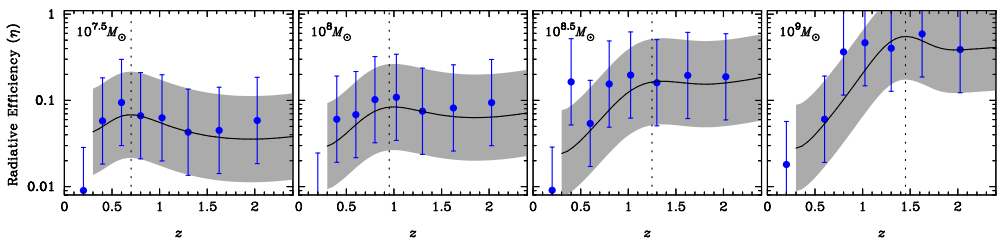} 
 \caption{Radiative efficiency evolution for different black hole mass. Solid lines and data points are
 the efficiency using the SMBH mass function derived from the galaxy LF and SMF, respectively.}
   \label{fig2}
\vspace*{-0.2 cm}
\end{figure}

The radiative efficiency evolution for different black hole masses is plotted 
in Fig.~\ref{fig2}.
Assuming that the radiative efficiency provides an effective indirect measure 
of the black hole spin, we propose that: (1) The accretion history of SMBHs and their spin evolution
can be characterized by two regimes: an initial phase of 
mass accumulation from prolonged accretion that spins up the hole, 
followed by a period of random, episodic accretion that spins down the hole 
toward lower redshifts (\cite[Wang et al. 2009]{Wang09}). 
(2) The evolution of the spin, like the global pattern of AGN activity, 
exhibits ``cosmic downsizing''. High-mass holes gain their masses earlier, reach the peak of their AGN activity 
earlier, and begin to spin down earlier.  Random accretion dominates their 
evolution below $z \approx 2$, whereas lower mass holes transition to this 
phase later, at $z \approx 1$ (\cite[Li et al 2012]{Li_etal12}).

\end{document}